\documentclass[]{interact}

\usepackage{epstopdf}
\usepackage[caption=false]{subfig}

\usepackage{natbib}
\bibpunct[, ]{(}{)}{;}{a}{}{,}
\renewcommand\bibfont{\fontsize{10}{12}\selectfont}
\usepackage[noperiod]{jabbrv}
\usepackage{url}
\usepackage{soul,threeparttable}
\usepackage{hyperref}
\theoremstyle{plain}

\theoremstyle{definition}

\theoremstyle{remark}

\begin{document}

\articletype{ARTICLE}

\title{Win Time In Favor of Treatment (WINFT) for Hierarchical Endpoints}

\author{
\name{Sahil S. Patel,\textsuperscript{a,b}\thanks{CONTACT Sahil S. Patel. Email: sspate27@ncsu.edu} Huiman Barnhart,\textsuperscript{b,c} Lu Mao,\textsuperscript{d} Roland A. Matsouaka,\textsuperscript{b,c} and Yuliya Lokhnygina\textsuperscript{b,c}}
\affil{\textsuperscript{a}Department of Statistics, North Carolina State University, Raleigh, North Carolina, USA; \\\textsuperscript{b}Duke Clinical Research Institute, Duke University, Durham, North Carolina, USA; \\\textsuperscript{c}Department of Biostatistics and Bioinformatics, Duke University, Durham, North Carolina, USA;\\\textsuperscript{d}Department of Biostatistics and Medical Informatics, School of Medicine and Public Health, University of Wisconsin, Madison, Wisconsin, USA.}
}

\maketitle

\begin{abstract}
 Standard win statistics methods determine a win, loss, or tie for a pair of subjects based on their worst outcomes (up to the end of study) that may not fully utilize all patients’ conditions or disease experience throughout the follow-up period. While the newly developed win-time statistics fully utilize all patients' longitudinal information, these statistics have been limited to time-to-event endpoints and require monotonic pattern of the events. As such, they are not applicable to any type nor number of hierarchical longitudinal endpoints. 
 
 We propose the {\it win time in favor of treatment} (WINFT), a general measure for any hierarchical longitudinal endpoints, that summarizes the total time a subject in the treatment group spends in a more favorable health state than a subject in the control group. Unlike existing win time methods, the WINFT does not require the component outcomes to be monotone and does not rely on modeling assumptions for estimating state probabilities. This flexibility allows analysis of a complex and diverse set of endpoints, and includes existing win time methods as special cases. Moreover, the WINFT is estimated based on U-statistics, which provide direct framework for variance estimation and confidence interval derivation, under independent censoring and missing at random assumptions, without expensive bootstrapping. We examine the performance of the proposed WINFT estimation method through simulation studies, and illustrate the method using data from the ACTT-1 COVID-19 and HF-ACTION trials. Overall, the WINFT offers a flexible and interpretable estimand for assessing treatment in clinical trial data with complex longitudinal outcomes.
\end{abstract}


\section{Introduction}\label{sec1}

Clinical trials often investigate multiple endpoints separately by selecting one primary endpoint and several secondary endpoints to assess the efficacy of an active treatment. Win statistics provide a framework for combining multiple of these endpoints in a prespecified hierarchy as the primary endpoint. This allows patient information on mixed types of endpoints, like survival and quality of life, to be analyzed simultaneously, while utilizing their relative level of importance across the endpoints experienced by patients. By comparing endpoints between a pair of patients (one from the treatment arm and one from the control arm) along the hierarchy, a winner, loser, or a tie can be established in terms of which participant had a more favorable outcome. This hierarchical endpoint analysis is the basis of estimands like the win ratio, win odds, net benefit, and the the desirability of outcome ranking (DOOR) \citep{Pocock_Ariti_Collier_Wang_2012, Buyse_2010, dong2020win, brunner2021win, barnhart2025sample}. 

A major drawback with the standard win statistic methods, however, is the aggregation of longitudinal information where a win, loss, or a tie for each pair is determined based on the worst status of each patient's outcome. This neglects or overlooks information accumulated throughout the patients' disease journey, i.e., what happened to the patients before they experienced their worst outcomes during the trial. 
\begin{figure}[h]
    \centering
    \caption{Example trial with 5 participants per arm, followed for 36 months. }
    \label{fig:intro_trajectory}
    \includegraphics[trim={0.2cm 0.1cm 1cm 0.1cm},clip, width=0.95\linewidth]{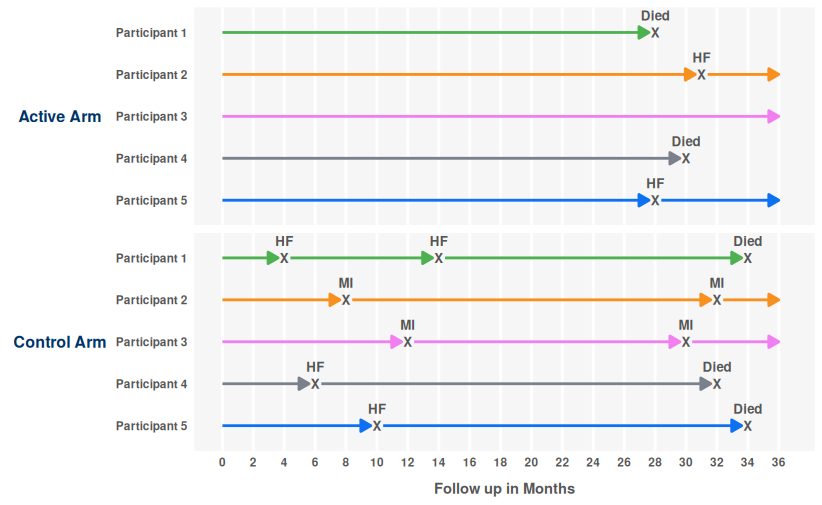}
    \begin{tablenotes}\tiny		
				\item	
				There are three endpoints, death, heart failure (HF) hospitalization, myocardial infarction (MI), ranked in that order. 
                \item The standard net benefit (of treatment vs. control) yields -0.12, which suggests the control treatment is better despite the numerous events experienced by control patients.
			\end{tablenotes}
\end{figure}

To illustrate why it is important to consider longitudinal information, let us consider a toy example. Suppose that 10 patients (with 5 randomized to each arm) were followed over 36 months. The patients were monitored for three hierarchical events, death, heart failure (HF) hospitalization, and myocardial infarction (MI), in this order, as shown in Figure \ref{fig:intro_trajectory}, with death as the most important event. 
If we use standard win statistics methods, we have a total of 25 pairwise comparisons (active vs. control arms), leading to 11 wins, 14 losses, and no ties in the active arm. Thus, the net benefit \citep{Buyse_2010} of the active arm over the control arm, i.e., the difference between the probabilities of wins and probabilities of losses, is -0.12. This means that the control treatment is favored over the active treatment when only looking at the worst states of the patients. However, it is clear that the control arm has more events than the active arm early in the study. The control arm is favored because the net benefit (as all standard win statistic methods) focuses on the worst states of patients: these worst states occurred slightly later in the control than in the active arm while the other events that occurred within the first  16 months (mainly among control patients) were ignored. 
Hence, win statistics do not always tell the full story when comparing the treatment and control arms. 

To incorporate the full longitudinal information, \cite{Mao_2023}, \cite{troendle_use_2024}, and others \citep{xu_novel_2024, rohde_bayesian_2024, shu_longitudinal_2025} have proposed win time statistics. For each pair of patients, the win time statistics do not focus solely on the assessment of worst endpoints over the common follow up time; they also account for each patient's clinical experience at every observational time point. Thus, for a pair of patients, the win time statistics evaluate and calculate the length of time each patient spends in better clinical states compared to the other patient.  Unlike the standard win statistics that only consider worse outcomes and give one verdict of win, loss, or tie for each pair of patients, the win time statistics consider a patient's entire clinical experience over the study period and allows the assessment of win, loss, or tie at different specific time points. For example, in Figure \ref{fig:intro_trajectory}, since each patient's status is recorded each month for 36 months, there would be 36 verdicts of wins, losses, or ties to consider when comparing pairs of active vs. control patients.

Mao developed the {\it restricted mean time in favor of treatment} (RMT-IF) to compare how well treatment and control arms work over a prespecified period of time (a "restricted" time window) \citep{Mao_2023}. The RMT-IF quantifies the cumulative net benefit by measuring the expected net time a patient in the active arm spent in a better clinical state (lived longer or experienced fewer bad events) than a control patient over the time window. 
The estimation of RMT-IF is formed by functions of state probabilities for each of the hierarchical events at a given time. The state probabilities are estimated by Kaplan-Meier (KM) survival functions of time to a specific type of event or higher order of such event. In the toy example, with death, HF hospitalization, and MI as hierarchical endpoints (see Figure \ref{fig:intro_trajectory}), the KM curves for each group are computed for (1) time to death, (2) time to death or HF, and (3) time to death, HF or MI. The RMT-IF requires all hierarchical states to be defined monotonically over time. If a patient has a HF hospitalization, which is ranked higher than a myocardial infarction (MI), then the HF hospitalization status is carried forward and overrides any MI that may occur after, as the patient's condition had already worsened to the “HF hospitalization state.” Furthermore, RMT-IF only works for time-to-event endpoints; thus, other types of endpoints, such as binary or quality-of-life (continuous) endpoints, cannot be used when we want to evaluate the RMT-IF. 

\cite{troendle_use_2024} defined several win time statistics for time-to-event hierarchical endpoints to measure the total health benefits of a treatment. Since, in this paper, we focus on absolute measures, their most relevant win time statistics are the {\it expected win time} (EWT) and the {\it pairwise win time} (PWT). Similar to the RMT-IF \citep{Mao_2023}, both EWT and PWT also look into the whole patient's disease experience to compare pairs of patients (one treated, one control) over a given period and require monotonic clinical states. The ``win time" is the extra time the treated patient survives or remains healthier than the control patient. The EWT is the average difference in these ``win times" between treated and control patients in the study through the effective follow-up time (usually the minimum of follow-up times in the treatment and control groups). 
If the effective follow-up time is the same as the chosen restricted time used in RMT-IF, then both EWT and RMT-IF are estimating the same exact quantity, with the core difference lying in the method of estimation of the state probabilities. \cite{troendle_use_2024} used both Kaplan-Meier curves and Markov-based approaches to estimate probabilities of being in various clinical states through time.

The PWT estimates the net amount of time a treated patient is in a more favorable state compared to a patient in the control arm over the average effective follow-up time. However, unlike EWT and RMT-IF that estimate state probabilities, the PWT uses pairwise comparisons for estimation. Specifically, for each pair of subjects (one treated and one control), paired win time difference is first defined as the sum of the win functions (defined as 1 for win, -1 for loss and 0 for tie for the treatment subject) at each time point over the pair's common follow-up time, and then PWT is defined as the average of all possible paired win time differences. If all subjects have complete follow up over a fixed period, for example, 36 months in Figure \ref{fig:intro_trajectory}, then the PWT is equal to the sum of net benefit values calculated at each of the 36 months. Equivalently, each pairwise comparison has 36 determinations of win, loss or tie at every month, rather than a single determination of win, loss, or tie in the standard win statistics framework. Therefore, when all subjects have complete follow up over the same fixed time period, the RMT-IF, EWT and PWT are essentially estimating the same quantity, although using different estimation methods. 

While the above win time statistics have been developed only for time-to-event endpoints, another estimand, the {\it days of benefit} (DOB), was developed for a single ordinal endpoint that is not necessarily a time-to-event endpoint. Days of benefit is defined as the sum of the net benefit of treatment over control based on the ordinal endpoint at each day over a pre-specified follow-up period. Unlike the above win time statistics (RMT-IF, EWT, and PWT), the DOB does not require monotonicity, i.e., a patient's status can get worse or better at any time. For example, if the ordinal endpoint has five levels: 1 = no symptom, 2 = mild symptoms, 3 = moderate symptoms, 4 = severe symptoms, 5 = death, then the patient's condition can go up from mild symptoms to severe symptoms and down to moderate symptom over a three day time span. \cite{rohde_bayesian_2024} developed a Bayesian procedure to analyze such an ordinal endpoint collected longitudinally by assuming a Markov structure. 

Although the existing win time statistics deal with longitudinal hierarchical endpoints, they are limited to time-to-event endpoints with monotonic restriction or to one ordinal endpoint. However, clinical trials often record longitudinal endpoints of other types to evaluate treatments and predict clinical benefits on how a patients feels, function, or survives \citep{detmar2002health,ponikowski2018esc}. Therefore, there is a need to develop win-time-like estimand that can handle all types of endpoints without restriction. In this paper, we propose the {\it win time in favor of treatment} (WINFT), which accommodates any type and any number of hierarchical endpoints collected longitudinally. The WINFT includes the RMT-IF, EWT, PWT and DOB as special cases. Furthermore, the WINFT does not require any monotonicity of the disease progression and its estimator is robust under the assumptions of independent censoring and missingness completely at random (MCAR). 

The rest of the paper is organized as follows. In Section \ref{sec2:methods}, we describe the proposed WINFT measure for longitudinal hierarchical endpoints and illustrate its use on examples. Then, we provide inference methods for the WINFT. In Section \ref{sec3:simulations}, we summarize the results of the simulation studies to evaluate the performance of the proposed estimation method, while in Section \ref{sec4:data_analysis} we consider two illustrative examples. Finally, in Section \ref{sec5:discussion} we discuss and draw the conclusions from the above results.

\section{Methods}\label{sec2:methods}
\subsection{Win Time in Favor of Treatment}
\subsubsection{Notation and Definition}
Suppose we want to compare the effects of two treatments, A (active) and B (control) on a set of endpoints over a pre-specified time period. Let $\bm{Y}:=(Y_1, \dots, Y_K)$ be the hierarchy of prespecified endpoints under study, ranked in descending order by clinical importance or disease severity from the most important (or most severe) $Y_1$ to the least important (or least severe) endpoint $Y_K$ measured longitudinally. The endpoints $Y_k$,  $ k=1, \dots, K$, can be of mixed types, such as binary, ordinal, continuous, or counts endpoints. 

Suppose there are $S$ unique observation times, $t_1 < \dots <t_S$ over a study follow-up time $\tau=t_S$. Let $\bm{Y}_{t_j}:=(Y_{t_j,1}, \dots, Y_{t_j,K})$ be the hierarchical endpoints available at time $t_j$, also ranked by clinical importance or disease severity. Based on $\bm{Y}_{t_j}$, we write $\bm{Y}^A_{t_j} \succ \bm{Y}^B_{t_j}$ to indicate that $A$ wins (over B) at time $t_j$, while $\bm{Y}^A_{t_j} \prec \bm{Y}^B_{t_j}$ means $A$ loses to B. Win or loss is determined by comparing pairs of patients (A vs. B),  starting with $Y_{t_j,1}$, to evaluate whether the patient in group A fares better (win) or worse (loss). If the comparison on $Y_{t_j,1}$ is a tie, then the pair is compared on $Y_{t_j,2}$. This process continues sequentially through the ordered endpoints in $\bm{Y}_{t_j}$, until there is a win (or loss) or we reach the last endpoint $Y_{t_j,K}$ and the comparison ends in a tie \citep{Pocock_Ariti_Collier_Wang_2012}. Note that for a time-to-event endpoint the comparison at time $t_j$ is based on the binary indicator of whether the patients experienced the endpoint.
 
We define the net benefit  $\text{NB}\big(\bm{Y}^A_{t_j}, \bm{Y}^B_{t_j}\big)$ at time $t_j$ as the difference between the probabilities of wins and losses in the treatment group  $A$, i.e.,
\begin{align}\label{netbenefit}
    \text{NB}\big(\bm{Y}^A_{t_j}, \bm{Y}^B_{t_j}\big) := P\big(\bm{Y}^A_{t_j} \succ \bm{Y}^B_{t_j}\big) - P\big(\bm{Y}^A_{t_j} \prec \bm{Y}^B_{t_j}\big)
\end{align}
The net benefit \eqref{netbenefit} is a snapshot of the treatment vs. control comparison at $t_j$ regardless of what happened at prior time points, except that death status carries forward. Thus, as the cumulative win time of the treatment over the control, the {\it win time in favor of treatment} (WINFT) is defined by the sum of net benefits over each observation time, i.e.,
\begin{align}\label{winft}
\text{WINFT}\big(\bm{Y}^A,\bm{Y}^B\big) := \sum_{j=1}^S \text{NB}\big(\bm{Y}^A_{t_j}, \bm{Y}^B_{t_j}\big).
\end{align}
The estimand $\text{WINFT}\big(\bm{Y}^A,\bm{Y}^B\big)$ provides a longitudinal perspective of the expected cumulative time a randomly selected patient in treatment A is in a better state than a randomly selected control patient. If the time points $t_1, \dots, t_S$ are evenly spaced with a unit such as day, month, or year, then WINFT can be interpreted in terms of this time unit. For example, in the toy example (Section \ref{sec1}, Figure \ref{fig:intro_trajectory}), WINFT is the expected number of months a randomly selected patient in the active treatment arm is in a better state than a randomly selected patient in control arm.

\subsubsection{Concept illustration}
To illustrate the concept of WINFT, we consider two specific examples. The first is the toy example of Section \ref{sec1} (illustrated by Figure \ref{fig:intro_trajectory}), where time points $t_1 = 1$ month, \dots, $t_S =36$ months are evenly spaced. The net benefits at each of these time points are derived using only the information available at this time point (with death status carried forward). Hence, each month, 25 pairwise comparisons are evaluated as treatment wins, losses, or ties. If two patients died at a time $t_j$, $j=1, \dots, S$, we consider their comparison a tie. 

We assume monotonicity of the event hierarchy: a patient who had a HF remains in the status of HF until (and if) death occurs, ignoring any MI events occurring after HF. With this monotonicity, the active arm dominates in the first 28 months and then has multiple losses in the last 8 (out of the 36 months), leading to a net win of 20.52 months, i.e., estimated WINFT = 20.52 using the estimation procedure described later. By contrast, the net benefit is -0.12, if evaluated over the 36 months without using all longitudinal data. The value $\text{WINFT}=20.52$ implies that the active arm has a net gain of 20.52 months compared to the control arm over the 36 month period. In practice, the monotonicity of the events should be defined based on clinical perspective and patient's preference. 

For the second example, we consider a study of a deadly disorder, with death and quality-of-life outcomes as the hierarchical endpoints. The follow-up times are $t_1 = 1$ week, $t_2=$ 1 month, $t_3=$ 4 months, and $t_5=$ 1 year, with the corresponding net benefits of 0.2, 0.5, 0.1, and 0. The overall net benefit at the conclusion of the study is 0, while the WINFT is 0.8. This WINFT can be interpreted as the cumulative net win time in favor of the treatment over 4 time points: 1 week, 1 month, 4 months, and 1 year. The interpretation of WINFT in this scenario is more nuanced due to unequal time periods. Our focus henceforth will be on equally spaced follow-up times where WINFT has a simple and straightforward interpretation in terms of the length of the time, rather than  the number of follow-up times. 

\subsection{Examples of WINFT}\label{compare}
\subsubsection{Existing win time statistics as special cases of WINFT}
To show how the WINFT fits in the existing literature, we show that it is a generalization of existing estimands. We start with the conditions under which the WINFT is equivalent to the life expectancy difference (LED) and the difference in restricted mean survival time (RMST) between two treatment groups \citep{Uno_Claggett_Tian_Inoue_Gallo_Miyata_Schrag_Takeuchi_Uyama_Zhao_et}. Then, we establish that WINFT is a generalization, and a more flexible version, of the restricted mean time in favor of treatment (RMT-IF) \citep{Mao_2023}, the expected win time (EWT), and pairwise win time (PWT) \citep{troendle_use_2024}.

\textit{Life expectancy difference (LED)}. Suppose we only have a single time-to-event endpoint, recorded in number of days, for a study duration of $T$ days. The LED between the treatment groups is defined as the number of days a randomly selected treated individual is expected to live longer (or in a better state) than an individual in the control group \citep{Uno_Claggett_Tian_Inoue_Gallo_Miyata_Schrag_Takeuchi_Uyama_Zhao_et}. Note that at each day a time-to-event endpoint is equivalent to a binary indicator with values equal to 0 (no event) or 1 (an event). A patient starts in state 0 until an event occurs, at which point he transitions to state 1. He stays in state 1 thereafter until $T$ (when monotonicity applies) or may return to 0 (in absence of monotonicity). When monotonicity applies (e.g., with time-to-death endpoint) the WINFT based on this binary indicator over the time period of $T$ days is the number of days of benefit of the treatment over the control, i.e., the number of days the treatment group is expected to live longer than the control, which is exactly the LED.

\textit{RMT-IF and EWT as special cases of WINFT.} 
A core assumption of RMT-IF and EWT is the monotonicity of the state spaces, where monotonicity means that once a higher-ranked event occurs, any lower-ranked information is forfeited. Both RMT-IF and EWT are defined for hierarchical time-to-event endpoints under the monotonicity assumption. If we convert each time-to-event endpoint into a longitudinal endpoint (i.e. with binary indicators at each prespecified time points  $t_1, \dots, t_S$) and assume monotonicity, then both RMT-IF and EWT are equivalent to the WINFT over the restricted time interval $[0,t_S]$. 

To illustrate this equivalence, we consider the following example where death, stroke, and myocardial infarction (MI) are three hierarchical events, ranked from highest to lowest. Let $\text{death}_t,$ $\text{stroke}_t,$ $\text{MI}_t$ denote, respectively, the binary indicators of death, stroke, and MI at time $t$. For $t=1$, define $Y_{1,1}=\text{death}_1$, $Y_{1,2} = \text{stroke}_1,$ $Y_{1,3} = \text{MI}_1$. At time $t>1$, let us define $Y_{t,1} = \max\!\left(Y_{(t-1),1},\text{death}_t\right),$ $Y_{t,2} = \max\!\left(Y_{(t-1),2},Y_{(t-1),1},\text{stroke}_t\right),$ and $Y_{t,3} = \max\!\left(Y_{(t-1),3}, Y_{(t-1),2},Y_{(t-1),1},\text{MI}_t\right)$. In this formulation, the higher ordered event in preceding time replaces the lower ordered event at the current time to ensure the monotonicity. Then the WINFT based on the longitudinal binary indicators of $(Y_{t,1}, Y_{t,2}, Y_{t,3})$ is the same as RMT-IF and EWT based on time to death, time to stroke and time to MI over restricted time of $[0,t_S]$.

\textit{In special circumstances, WINFT is also the pairwise win time (PWT) \citep{troendle_use_2024} and the days benefit \citep{rohde_bayesian_2024}}. By using restricted time interval $[0, t_S]$, the PWT estimates the EWT by pairwise comparisons and thus is a special case of WINFT. As shown in Section \ref{estimation}, both the estimators of WINFT and PWT are based on pairwise comparisons and are identical under restricted time interval, in absence of censoring or missing data. Finally, we can also draw a parallel between the {\it days of benefit} of \cite{rohde_bayesian_2024} and the WINFT, since the days of benefit is equivalent to the WINFT for a single longitudinal ordinal endpoint.  

Overall, we demonstrated that by redefining events or making certain assumptions on the state space, the WINFT matches existing longitudinal metrics for hierarchical endpoints. A key difference between WINFT and days of benefit, RMT-IF, EWT, and even RMST is its generalizability, as WINFT can be defined for a wide class of endpoints. By contrast, RMST is only applicable to a single time-to-event endpoint. RMT-IF (and the closely related EWT), expand RMST to accommodate hierarchical endpoints, by requiring the endpoints to be monotonic. The days of benefit allows non-monotonic states, but only is only applicable in a scenario with a single longitudinal ordinal endpoint or, equivalently, multiple binary longitudinal endpoints. 

\subsubsection{WINFT as a generalization of the existing win-time statistics}

To see how WINFT extends beyond the existing methods, consider a scenario with death and recurrent HF hospitalizations. Let $[0,t_S]$ be the time interval of interest.  There may be two different ways to define hierarchical endpoints depending on the need for interpretation. If the number of HF hospitalizations is considered a better endpoint (than just occurrence of one or more HF hospitalizations) to capture the treatment effect, the two hierarchical endpoints at time $t$ can be defined as death and the cumulative number of HF hospitalizations by time $t$. However, if HF hospitalization as a binary endpoint of being in (1) or out (0) of hospital captures the treatment effect better than the number of HF hospitalizations, then the two hierarchical endpoints at time $t$ can be defined by death and HF hospitalization at that time. In this way, the length of the hospital stay is captured because the subject may be in the hospital at one time and then out of the hospital at a later time. Methods like RMT-IF or EWTR cannot accommodate the latter set of hierarchical endpoints due to the monotonic requirement, while number of hospitalizations satisfies the monotonicity assumption and can be handled by existing methods. By contrast, whether we consider death and the number of HF hospitalizations or simply death and HF hospitalization (i.e., being in or out of a hospital), the WINFT remains applicable.

\subsection{Estimation and Inference}\label{estimation}

Considering that WINFT is just the cumulative net benefit, inference methodology can be established from U-statistics \citep{bebu_large_2016}. To provide robust inference under independent censoring and missing data, we follow the derivations of \cite{dong_inverse-probability--censoring_2020}. Let there be $n_A$ treatment and $n_B$ control participants in the study, and let $t_1 < ... < t_S$ be the follow-up time points, where at least 1 treatment and 1 control participant are observed. Let $\bm{Y}_{i}^z(t_j)=(Y_{i,1}^z(t_j),Y_{i,2}^z(t_j),...,Y_{i,K}^z(t_j))$ be the observed hierarchical endpoints, at time $t_j$, of the $i$th individual in arm $z\in \{A, B\}$. The endpoints are ordered from the most important $Y_{i,1}^z(t_j)$, to the least important $Y_{i,K}^z(t_j)$. In the case where death is the first prioritized endpoint, we define $Y_{i,1}(t_j)$ as a binary indicator on whether the patient has died by time $t_j$. Specifically, $Y_{i,1}^z(t_j)=\mathbb{I}\{D_i^z\leq t_j\}$, where $D_i^z$ is the death time. We define the net benefit kernel for subject $i$ in group A and subject $l$ in group B as,
    \begin{align}
       W\big(\bm{Y}_i^A(t_j), \bm{Y}^B_l(t_j)\big)=
       \begin{cases}
        1 \text{ for } \bm{Y}^A_i(t_j) \succ \bm{Y}^B_l(t_j) \\
        -1 \text{ for } \bm{Y}^A_i(t_j) \prec \bm{Y}^B_l(t_j) \\
        0 \text{ for } \bm{Y}^A_i(t_j) =\bm{Y}^B_l(t_j) \\
        \end{cases} 
    \end{align}
where the kernel $W_{il}(t_j):= W\big(\bm{Y}_i^A(t_j), \bm{Y}^B_l(t_j)\big)$ can be interpreted as a 1-on-1 comparison between a treatment and control participants at time $t_j$. A treatment win is encoded as a 1, a control win as a -1, and a tie as a 0. A win is decided by looking at the outcomes successively. Starting with the most important outcome, $Y_{1}$, we compare whether $Y_{i,1}^A$ is better (or worse) than $Y_{l,1}^B$. If they are tied, we compare the subjects on the next endpoint $Y_{2}$. This process continues until a winner is decided or the comparison ends in a tie. If a patient has died or been censored, we do not consider any lower endpoints because the decision would either have been made based on the death endpoint or there is no information to compare the pair. If both the participants being compared have died, the kernel evaluates to 0. 

Since a person cannot be censored after death, we must account for bias in the estimation of WINFT beyond just evaluating the kernel as a two-sample $U$-statistic. We assume that each arm has an independent right-censoring time distribution, captured by the random variable $C^z$, and define $G^z(t)=P(C^z>t)$, the probability of not being censored by time $t$. We also consider a missing data process and define $O_i^z(t)$ the missingness indicator for non-death endpoints, where $O_i^z(t)=1$ if the endpoint $\bm{Y}_i^z(t_j)$ of $i$th person in group $z\in \{A, B\}$ is observed when they are alive and uncensored. We define $o^z(t)=P(O^z_i(t)=1)$ and we assume that any missing data is missing completely at random and that censoring is non-informative.

There are two scenarios for which an individual's status can be used in estimating the net benefit at time $t_j$. The first scenario happens when a participant has died before $t_j$ and their death was observed. The second scenario happens when we know the patient has not been right-censored yet (they are observed at some later time), has not died yet, and their information is not missing. We define the indicator variables $A^z_i(t_j)$ and $B^z_i(t_j)$ for both of these scenarios as
\begin{align*}
   A_i^z(t_j)&:=\mathbb{I}\{D_i^z<t_j, D_i^z<C_i^z\}=\mathbb{I}\{D_i^z<t_j\}\mathbb{I}\{D_i^z<C_i^z\}\\
B_i^z(t_j)&:=\mathbb{I}\{D_i^z>t_j, C_i^z>t_j, O_i^z(t_j)=1\}=\mathbb{I}\{D_i^z>t_j\}\mathbb{I}\{C_i^z>t_j\}O_i^z(t_j).
\end{align*}
Similar to \cite{dong_inverse-probability--censoring_2020}, we define weights $\Omega^z_i(t_j)$ for the $U$-statistic estimator to adjust for possible censoring or missingness and use them to derive the estimator $\widehat{\text{WINFT}}$ for WINFT,
\begin{align*}
\Omega_i^z(t_j)&:=\frac{A_i^z(t_j)}{G^z(D_i^z)}+\frac{B_i^z(t_j)}{G^z(t_j)o^z(t_j)}\\
\widehat{\text{WINFT}}&=\frac{1}{n_An_B}\sum_{i=1}^{n_A}\sum_{l=1}^{n_B}\sum_{j=1}^{S}\Omega_i^A(t_j)\Omega_l^B(t_j)W_{il}(t_j)
\end{align*}
However, if death is not a relevant endpoint for the analysis, then $\Omega_i^z(t_j)$ reduces to $\frac{B_i^z(t_j)}{G^z(t_j)o^z(t_j)}$ as we no longer need to handle the persistent observed death indicator. Moreover if missing data or censoring is not an issue, the $\text{WINFT}$ just reduces to the PWT. Assuming $G^z(t)$ and $o^z(t)$ are known, then $\widehat{\text{WINFT}}$ is just a two-sample $U$-statistic, 
$$\widehat{\text{WINFT}}=\frac{1}{n_An_B}\sum_{i=1}^{n_A}\sum_{l=1}^{n_B}f\big(\bm{Y}_i^A, \bm{Y}^B_l\big)$$
with $f\big(\bm{Y}_i^A, \bm{Y}^B_l\big)=\displaystyle \sum_{j=1}^{S}\Omega_i^A(t_j)\Omega_l^B(t_j)W_{il}(t_j)$ the kernel of the two-sample $U$-statistic.
Therefore, all available results about the unbiasedness and asymptotic normality of $\widehat{\text{WINFT}}$ are given quite easily since the kernel itself is unbiased (see Appendix \ref{appA}). The estimator of the variance of $\widehat{\text{WINFT}}$ follows from the Hajek projection principle,

 \begin{align*}
    \widehat{V}(\widehat{\text{WINFT}}) :&= \frac{1}{n_A^2}\sum_{i=1}^{n_A}\phi_i^2 + \frac{1}{n_B^2}\sum_{k=1}^{n_B}\psi_k^2\\
   \text{ with }~ \phi_i :&= \frac{1}{n_B}\sum_{k=1}^{n_B} f(\bm{Y}^A_{i}, \bm{Y}^B_{k}) - \widehat{\text{WINFT}}\\\text{ and }~~
    \psi_k :&= \frac{1}{n_A}\sum_{i=1}^{n_A} f(\bm{Y}^A_{i}, \bm{Y}^B_{k}) - \widehat{\text{WINFT}}
\end{align*}
More often, the censoring and missing data distributions are unknown; we must use estimates for $G^z(t)$ and $o^z(t)$. For $G^z(t)$, we use a Kaplan-Meier estimate for the censoring survival distribution for each treatment group. We define the estimate for $o^z(t)$ as,

$$\widehat{o}^z(t):= \sum_{i=1}^{n_z}\frac{\mathbb{I}\{D_i^z>t, C_i^z>t\}O_i^z}{\mathbb{I}\{D_i^z>t, C_i^z>t\}}$$

For all of the subsequent analysis, we use $\widehat o^z(t)$ and $\widehat G^z$ as estimates for $o^z(t)$ and $G^z$. While the variance estimate can be updated to handle using estimates for these distributions, like in \cite{ozenne_asymptotic_2021}, we see that this is an unnecessary complication at even rather small sample sizes. Overall, it is important to notice that the construction of WINFT is purely nonparametric via pairwise comparisons of individuals, similar to PWT. This is in contrast to existing methods like RMT-IF and EWT which estimate state probabilities. Moreover, we can directly estimate the variance instead of using a bootstrapping procedure as in EWT and PWT.

\section{Simulation Study}\label{sec3:simulations}

To evaluate the performance of the WINFT, we conducted two sets of simulation studies. The first set evaluates the accuracy of our estimation procedure for WINFT. By simulating data according to a known WINFT, we can directly investigate WINFT estimates under different sample sizes, censoring, and missingness levels to assess coverage properties. A key aspect of this simulation is the non-monotonic state space, which no existing method could handle appropriately, as discussed in Section \ref{compare}. 
In our second set of simulations, we compared WINFT to RMT-IF and EWT. By comparing power under different modeling conditions, we can evaluate how the more general construction of WINFT compares to existing models in estimating the win time and corresponding variance. 

Throughout the simulations, we expect the results to return WINFT point estimates around the true values of 0 (under the null hypothesis of no treatment difference) and 1.7 (under the alternative hypothesis) while the 95\% coverage probabilities are within the interval $[0.945, 0.954]$, based on the 10,000 simulated data. 

\subsection{Assessing WINFT Bias and Variance Estimation}
\label{sec:valid_WINFT}
To investigate the estimator of WINFT and its variance, we start by conducting a simple simulation study over an ordinal endpoint with 8 levels. Using the results from the Adaptive
Covid-19 Treatment Trial (ACTT-1), which
evaluated remdesivir against a placebo (reported in Table 2 of \cite{beigel_remdesivir_2020}), we posit a Markov model for the transition probabilities between states 1 through 8 for participants tracked over 15 days. The meanings of each state are presented in Table \ref{tab:scores}, where 1 and 2 correspond to being out of the hospital, 3 through 7 correspond to being in the hospital with increasing levels of severity, and 8 represents death.

\begin{table}[h]
\tbl{Interpretation of scores from the ACTT-1 clinical trial \citep{beigel_remdesivir_2020}.}
{\begin{tabular}{ll} 
\toprule
\textbf{Score} & \textbf{Meaning} \\
\midrule
1 & Not hospitalized, no limitation on activities \\
2 & Not hospitalized, imitation on activities and/or requiring home oxygen \\
3 & Hospitalized, not requiring supplemental oxygen, no longer requires ongoing medical care \\
4 & Hospitalized, not requiring supplemental oxygen, requiring ongoing medical care \\
5 & Hospitalized, requiring supplemental oxygen \\
6 & Hospitalized, on non-invasive ventilation or high flow oxygen devices \\
7 & Hospitalized, on invasive mechanical ventilation, or ECMO \\
8 & Death \\
\bottomrule
\end{tabular}}
\label{tab:scores}
\end{table}

After positing a Markov model with constant transition probabilities per day, we also assume that a participant only changes their score once. From these assumptions, we can derive the "true" transition probabilities for a Markov model using observed counts of participant statuses at baseline and on day 15. Further details are provided in Appendix \ref{appB}.  Based on these transition probabilities, we can simulate patient trajectories and calculate the true WINFT: 1.77 days benefit of the treatment over the control. We simulated 10000 trials with 100, 200, 400, and 1000 participants under a 1:1 randomization. We investigated the average of estimated WINFTs and compared the Monte Carlo (MC) variance, i.e., the variance of the 10,000 estimated WINFTs, to the average of the estimated variances from each estimated WINFT. We also investigated the performance of the WINFT under the null hypothesis of no treatment difference using the same treatment and control transition probabilities to simulate the data. Note that the patient states are not monotone since they can increase or decrease; both EWT and RMT-IF can not be used to assess treatment effect as they require monotone state transitions. However, WINFT can estimate the net number of days of benefit from the treatment.

The results, shown in Table \ref{tab:winft_variance}, indicate that overall under both the null and oracle models, WINFT is consistent for the true number of days benefited by the treatment over the control. Under all the sample sizes considered, the average estimated WINFT is approximately equal to the true value of WINFT. The average estimated variance and the Monte Carlo estimate of variance are also quite close, meaning that the variance estimation via the $U$-statistic is working well, even at low sample sizes. Lastly, we see that most of the coverage probabilities are well-calibrated and approximately hit the desired 95\% coverage level; as these coverage probabilities are expected to be within the interval $[0.945, 0.954]$, based on the 10,000 simulated data.

\begin{table}[h]
\tbl{WINFT Point Estimates, Variance Estimates, and Empirical 95\% Confidence Interval.}
{\begin{tabular}{lccccc} 
\toprule
\textbf{\shortstack{True\\$WINFT$}} &
\textbf{$n$} &
\textbf{\shortstack{Average\\$\widehat{WINFT}$}} &
\textbf{\shortstack{Average Variance\\Estimate}} &
\textbf{\shortstack{MC Variance\\Estimate}} &
\textbf{\shortstack{Empirical 95\%\\CI Coverage}} \\
\midrule
1.77 & 100  & 1.768 & 2.041 & 2.081 & 0.944 \\
1.77 & 200  & 1.758 & 1.028 & 1.036 & 0.946 \\
1.77 & 500  & 1.770 & 0.412 & 0.426 & 0.945 \\
1.77 & 1000 & 1.775 & 0.206 & 0.213 & 0.945 \\
\midrule
0 & 100  & -0.018 & 2.068 & 2.142 & 0.942 \\
0 & 200  & 0.000 & 1.042 & 1.028 & 0.950 \\
0 & 500  & 0.002  & 0.418 & 0.428 & 0.946 \\
0 & 1000 & 0.002  & 0.209 & 0.213 & 0.946 \\
\bottomrule
\end{tabular}}
    \begin{tablenotes}\tiny		
                 \item Coverage under the oracle derived probabilities (WINFT=1.77) and when there is no treatment effect (WINFT=0).
			\end{tablenotes}
\label{tab:winft_variance}
\end{table}

\subsection{Assessing WINFT Under Censoring}

To investigate how robust the WINFT is under censoring, we used the previous data generating processes (with a sample size of $n=500$) as in Section \ref{sec:valid_WINFT}, but added an independent exponential censoring process over it. We considered 3 censoring scenarios where the expected censored participants by the end of the study were 10\%, 20\%, and 30\%, respectively.

Based on the results in Table \ref{tab:winft_cens}, we see that WINFT is robust to independent censoring. Even when 40\% of the population is expected to be censored by the end of the study, the coverage of the 95\% confidence interval is near nominal 0.95 level and the estimates for variance closely match the Monte Carlo variance estimates.

\begin{table}[ht]
\tbl{WINFT Point Estimates, Variance Estimates, and Empirical 95\% Confidence Interval.}
{\begin{tabular}{lccccc} 
\toprule
\textbf{\shortstack{True\\WINFT}} &
\textbf{\shortstack{Expected\\Censoring (\%)}} &
\textbf{\shortstack{Average\\$\widehat{WINFT}$}} &
\textbf{\shortstack{Average Variance\\Estimate}} &
\textbf{\shortstack{MC Variance\\Estimate}} &
\textbf{\shortstack{Empirical 95\%\\CI Coverage}} \\
\midrule
1.77 & 10  & 1.765 & 0.425 & 0.429 & 0.948 \\
1.77 & 20  & 1.770 & 0.441 & 0.431 & 0.953 \\
1.77 & 40  & 1.771 & 0.482 & 0.465 & 0.953 \\
\midrule
0 & 10  & -0.004 & 0.431 & 0.437 & 0.951 \\
0 & 20  & -0.002 & 0.446 & 0.440 & 0.950 \\
0 & 40  & 0.004 & 0.487 & 0.478 & 0.955 \\
\bottomrule
\end{tabular}}
    \begin{tablenotes}\tiny		
                 \item Coverage under the oracle derived probabilities (WINFT=1.77) and when there is no treatment effect (WINFT=0).
                 \item A fixed sample size of $n=500$ was used.
			\end{tablenotes}
\label{tab:winft_cens}
\end{table}

\subsection{Assessing WINFT Under Missingness and Censoring}
Lastly, to investigate how robust WINFT is under missingness and censoring, we introduce an independent missingess process over the data generating procedure. We considered the probability of being missing on any given day equal to $p=10\%, 20\%$ and 30\%. Since the death date is typically known, we assume that the day a simulated participant dies the data cannot be missing.

The results, displayed in  Table \ref{tab:winft_missing}, show that under both censoring and missingness, WINFT remains robust. The average estimate across each level of missingness confirms that WINFT is consistent for the number of days benefit of the treatment over the control. The estimation of variance through $U$-statistic also appears robust as the average variance estimate closely matches the Monte Carlo variance estimate. Finally, inference through Wald confidence intervals appears to be valid as the 95\% confidence intervals contain the true WINFT at nearly the nominal level of 0.95.

\begin{table}[h]
\tbl{WINFT Point Estimates, Variance Estimates, and Empirical 95\% Confidence Interval.}
{\begin{tabular}{lccccc} 
\toprule
\textbf{\shortstack{True\\\text{WINFT}}} &
\textbf{\shortstack{Probability of\\\text{missing data}}}&
\textbf{\shortstack{Average\\$\widehat{\text{WINFT}}$}} &
\textbf{\shortstack{Average Estimated\\Variance}} &
\textbf{\shortstack{MC Variance\\Estimate}} &
\textbf{\shortstack{Empirical 95\%\\CI Coverage}} \\
\midrule
1.77 & 0.1  & 1.764 & 0.444 & 0.443 & 0.947 \\
1.77 & 0.2  & 1.757 & 0.447 & 0.434 & 0.954 \\
1.77 & 0.3  & 1.772 & 0.452 & 0.451 & 0.949 \\
\midrule
0 & 0.1  & -0.005 & 0.449 & 0.455 & 0.948 \\
0 & 0.2  & -0.008 & 0.453 & 0.444 & 0.954 \\
0 & 0.3  & 0.002  & 0.458 & 0.479 & 0.943 \\
\bottomrule
\end{tabular}}
    \begin{tablenotes}\tiny		
                 \item Coverage under the oracle derived probabilities (WINFT=1.77) and when there is no treatment effect (WINFT=0).
                 \item A fixed sample size of $n=500$ was used. Expected percent of participants censored by the end of the study = 20\%.
			\end{tablenotes}
\label{tab:winft_missing}
\end{table}

\subsection{Comparing WINFT against EWT, PWT, and RMT-IF}
When using monotone hierarchical time-to-event endpoints with a restricted time horizon, estimators of WINFT, EWT, PWT, and RMT-IF target the same estimand. The simulation studies here aim to assess unbiasedness and variability of the different estimation approaches and potential differential impacts due censoring. We compare estimation and inference for WINFT against EWT, PWT, and RMT-IF by simulating time-to-event data following the process of \cite{troendle_use_2024} (adapted from \cite{beyersmann_simulating_2009}) under a semi-competing risk scenario. In this simulation setup, death and censoring preclude any other events, while heart failure (HF) and myocardial infarction (MI) do not. The method uses a cause-specific hazard model for each treatment arm across all 4 events (MI, HF, death, censoring) where time to death, time to HF, and time to MI are the hierarchical endpoints in this order of importance. We assumed monotonicity of the events, a fixed time horizon of 1 year, and used a total sample size of $n=1000$. Further simulation details are provided in Appendix \ref{appE}. As we mentioned previously, time-to-event endpoints are equivalent to longitudinal binary endpoints, and we use the binary hierarchical endpoints longitudinally for estimation and inference of WINFT. 

 We investigated 3 scenarios, replicating each 2000 times. The first scenario, $\beta_1=\beta_2=\beta_3=0$, implies that there is no treatment effect. The second scenario, $\beta_1=\beta_2=\beta_3=-0.2,$ implies that the treatment impacts on each endpoint (MI, HF, and death) equally. The last scenario, $\beta_1=-0.1, \beta_2=-0.2,\beta_3=-0.3,$ represents the case where the treatment has the smallest impact on MI, a larger impact on HF, and the largest impact on death. We considered 3 levels of censoring: (1) no censoring; (2) censoring with $\alpha_4=-1$; (3) censoring with $\alpha_4=-0.7$. When $\beta_1=\beta_2=\beta_3=0$ and $\alpha_4=-1$ we expect approximately 26.82\% of all participants to be censored by the end of the 1-year follow-up. This increases to 34.21\% when $\alpha_4=-0.7$. 
 
 Overall, the expected censoring percentages (when there is a treatment effect) are given in the results Table \ref{tab:rr_estimands}. Estimation and inference on EWT and PWT were carried out using R-package, {\it wintime} \citep{troendle_wintime_2026}, while estimation and inference on RMT-IF was carried out with the R-package {\it rmt} \citep{mao_rmt_2021}.

\begin{figure}[h]
    \centering
    \caption{Distribution of EWT, PWT, RMT-IF, and WINFT when there is no treatment effect.}
        \includegraphics[trim={0.1cm 0.2cm 0.1cm 0.7cm},clip, width=0.9\linewidth]{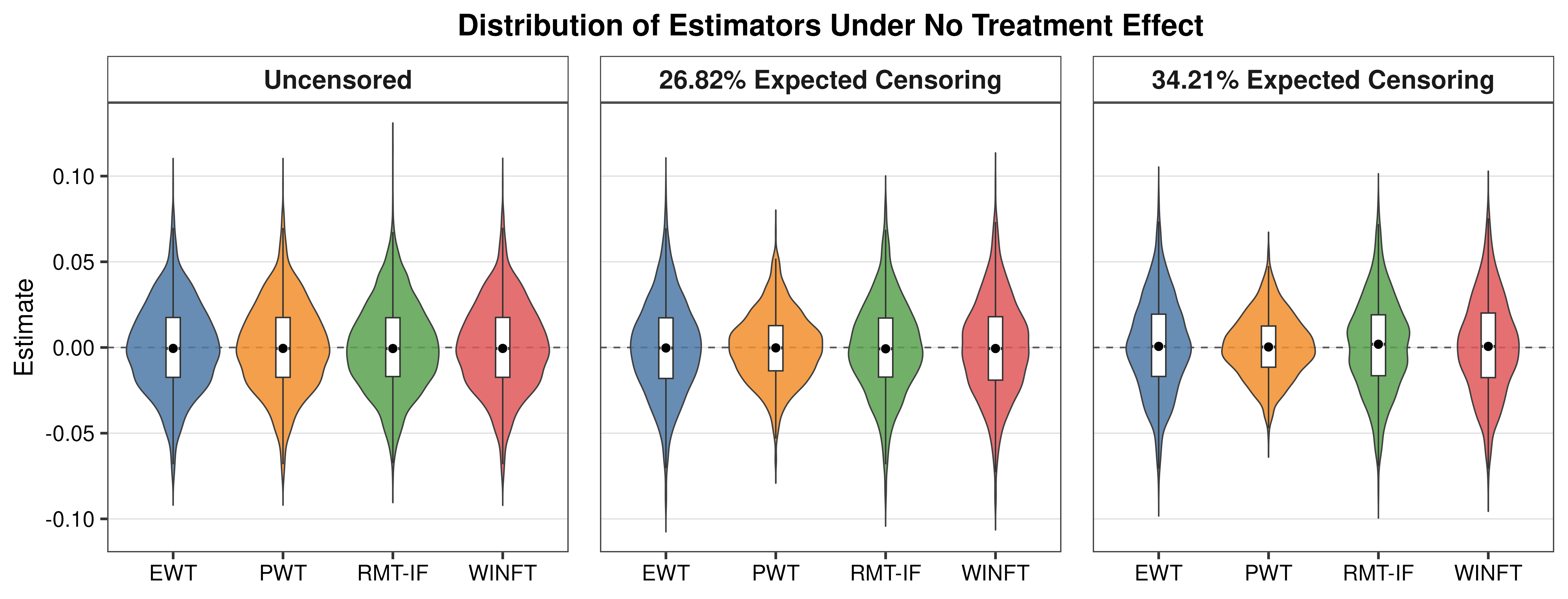}
      \begin{tablenotes}\tiny		
                 \item Different levels of censoring are introduced: the left most pane is uncensored and moving to the right increases censoring. 
                 \item $n=1000$ and $\beta_1=\beta_2=\beta_3=0$. When $\alpha_4=-1$, the overall expected censoring is 26.82\%, and 34.21\% when $\alpha_4=-0.7$
			\end{tablenotes}
    \label{fig:troendle_null}
\end{figure}

Figure \ref{fig:troendle_null} provides the violin plots on the distributions of all 4 estimators in the first scenario when there is no treatment effect. We can see that WINFT is consistent for the true value of 0. PWT decreases in variability as the level of censoring increases while all other estimators seem to increase in variability, as expected since PWT does not account for censoring.

\begin{figure}[h]
    \centering
    \caption{Distribution of EWT, PWT, RMT-IF, and WINFT when there is a treatment effect. 
    }
    \includegraphics[trim={0.1cm 0.2cm 0.1cm 0.7cm},clip, width=0.9\linewidth]{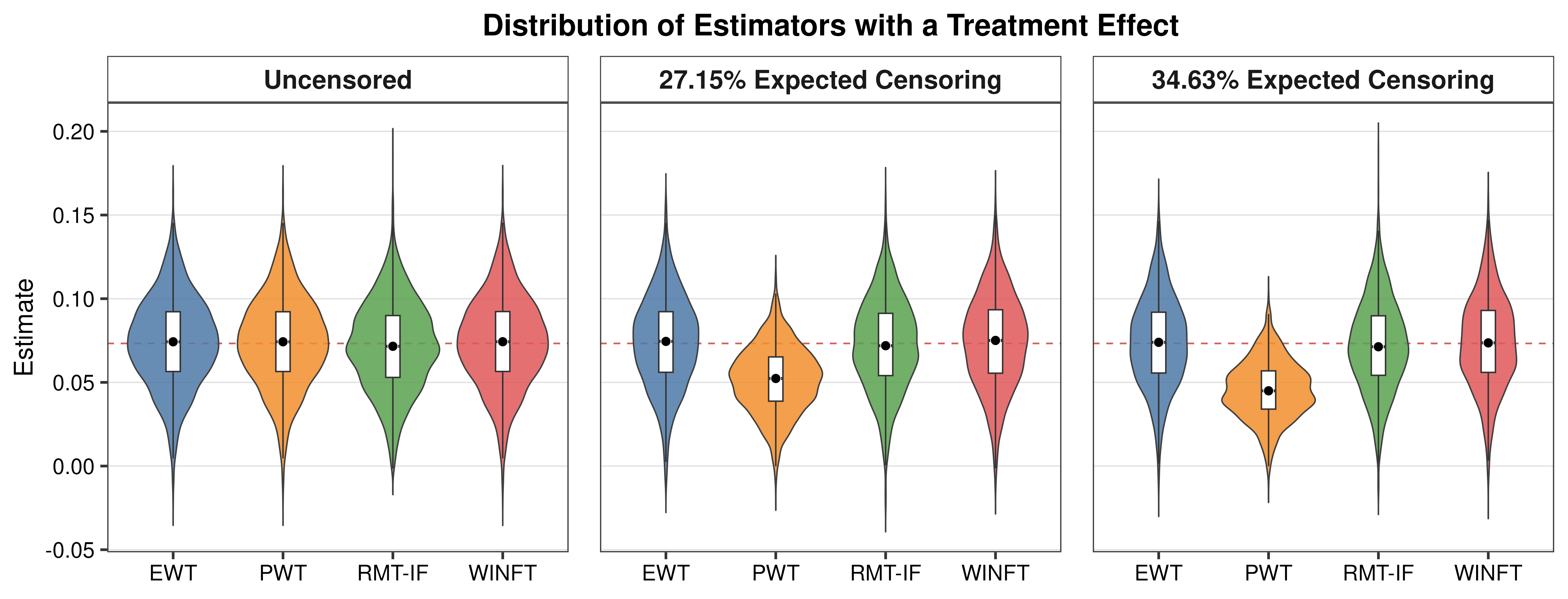}
    \begin{tablenotes}\tiny		
                 \item Different levels of censoring are introduced: the left most pane is uncensored and moving to the right increases censoring. 
                 \item $n=1000$ and $\beta_1=\beta_2=\beta_3=-0.2$ and the red dashed line represents the estimated true years benefit of the treatment over the control using the average of the EWT, RMT-IF, and WINFT simulations under no censoring. 
                 \item When $\alpha_4=-1$, there is an overall expected censoring of 27.15\%, and 34.63\% when $\alpha_4=-0.7$
			\end{tablenotes}
    \label{fig:troendle_alt}
\end{figure}

The simulated distribution of all 4 statistics in the second scenario, where there is a treatment effect, are displayed in Figure \ref{fig:troendle_alt}. We see that EWT, RMT-IF, and WINFT all estimate roughly the same value. While in the uncensored case, PWT is consistent for this same value, its estimator appears to be a biased  when censoring is present. This is due to the fact that PWT does not use weighting (or any other approach) in the pairwise comparisons, like WINFT, to correct for censoring. Consequently, we only investigated EWT and RMT-IF as comparators to WINFT when looking at rejection rates, i.e., power comparisons, as reported in Table \ref{tab:rr_estimands}. Similar to \cite{troendle_use_2024}, the hypothesis testing is the one-sided at 0.025 level at the null of no treatment effect. This means, under the null hypothesis, the rejection rates are expected to be within the interval $[1.82\%, 3.18\%]$, based on the 2,000 simulated datasets. 

\begin{table}[h]
\tbl{Rejection rates or power to reject the null of no treatment effect with 1-sided test at 0.025 level.}
{\begin{tabular}{cccccccc} 
\toprule 
\multicolumn{2}{c}{Censoring}&\multicolumn{3}{c}{Scenarios (\boldmath$\beta_1$, \boldmath$\beta_2$, \boldmath$\beta_3$)}&\multicolumn{3}{c}{Rejection rate (\%)}\\\cmidrule(lr){1-2}	\cmidrule(lr){3-5}	\cmidrule(lr){6-8}	
\textbf{Param. $\alpha_4$ }& \textbf{Expected (\%)} &
\shortstack{\boldmath$\beta_1$\textbf{(MI)}} &
\shortstack{\boldmath$\beta_2$\textbf{(HF)}} &
\shortstack{\boldmath$\beta_3$\textbf{(Death)}} &
\shortstack{\textbf{\boldmath{\text{EWT}}}} &
\shortstack{\textbf{\boldmath{\text{RMT-IF}}}}&
\shortstack{\textbf{\boldmath{\text{WINFT}}}} \\
\midrule
&0 & 0.0 & 0.0 & 0.0 & 2.20 & 2.95  & 2.5 \\
&0 & -0.2 & -0.2 & -0.2 & 83.15 & 83.65  & 83.05  \\
&0 & -0.1 & -0.2 & -0.3 & 86.70 & 84.75  & 86.75  \\
\midrule
& 26.82 & 0.0 & 0.0 & 0.0 & 3.05 & 2.40 & 2.80 \\
$-1$ & 27.15 & -0.2 & -0.2 & -0.2 & 76.95 & 77.70  & 76.00  \\
& 27.29 & -0.1 & -0.2 & -0.3 & 80.90 & 81.75  & 80.15  \\
\midrule
& 34.21 & 0.0 & 0.0 & 0.0 & 2.25 & 2.65  & 2.55 \\
$-0.7$& 34.63 & -0.2 & -0.2 & -0.2 & 76.95 & 76.60  & 75.45  \\
& 34.81 & -0.1 & -0.2 & -0.3 & 81.25 & 80.25  & 78.65 \\
\bottomrule
\end{tabular}}
   \begin{tablenotes}\tiny		
                 \item Calculated with $n=1000$, and $200$ bootstrap replications for the variance estimation of EWT.
                 \item The censoring parameter $\alpha_4$ is defined in the Appendix \ref{appE}.
	\end{tablenotes}
\label{tab:rr_estimands}
\end{table}

From Table \ref{tab:rr_estimands}, we see that the WINFT is slightly more conservative than the other two methods when there is truly a treatment effect, except in the 3rd scenario. The difference in rejection rates between RMT-IF, EWT, and WINFT differ by at most 2.6\% in the 3rd scenario with $\alpha_4=-0.7$. This suggests in scenarios with a treatment effect (i.e., 2nd and 3rd scenarios), WINFT tends to be slightly more conservative. Furthermore, under the null hypothesis, all the rejection rates are within $[1.82\%, 3.18\%]$, meaning the type-1 error is controlled reasonably well. 

Overall, despite being model free, the WINFT maintains the type-1 error control and has a power comparable to the power of the existing methods.

\section{Examples}\label{sec4:data_analysis}

To illustrate WINFT, we analyze data from two clinical trials: the ACTT-1 trial data with one ordinal endpoint and a subset of the HF ACTION trial data with two endpoints, i.e., time to death and number of heart failure hospitalizations.

\subsection{ACTT-1 Trial}

The Adaptive Covid-19 Treatment Trial (ACTT-1) was a double-blind randomized controlled trial to evaluate the efficacy of remdesivir for COVID-19 treatment \citep{beigel_remdesivir_2020}. The study enrolled 1062 patients hospitalized with a COVID-19 infection (and a severity score between 4 and 7, see Table \ref{tab:scores}). The patients were stratified by study site and severity score and then randomized to either remdesivir or placebo. The primary outcome was the time to recovery and one of the secondary outcomes was the 8-level ordinal severity score at day 15. Since this ordinal outcome was measured longitudinally over 28 days, we use it to illustrate the WINFT methodology. A patient's status on this ordinal scale can go up or down in severity on any given day. 

 The data we obtained did not contain information on limitation of activities or home oxygen need, when a patient was not hospitalized. Therefore, we combined the first two severity-score levels to have a 7-level ordinal endpoint. Scores for a patient were missing once the patient was lost to follow-up, e.g., discharged, or until the patient was hospitalized again. 

 \begin{figure}[h]
    \centering
    \caption{Distribution of ACTT-1 scores per day by treatment arm.}
        \includegraphics[trim={0.1cm 0cm 0.1cm 0cm},clip, width=1\linewidth]{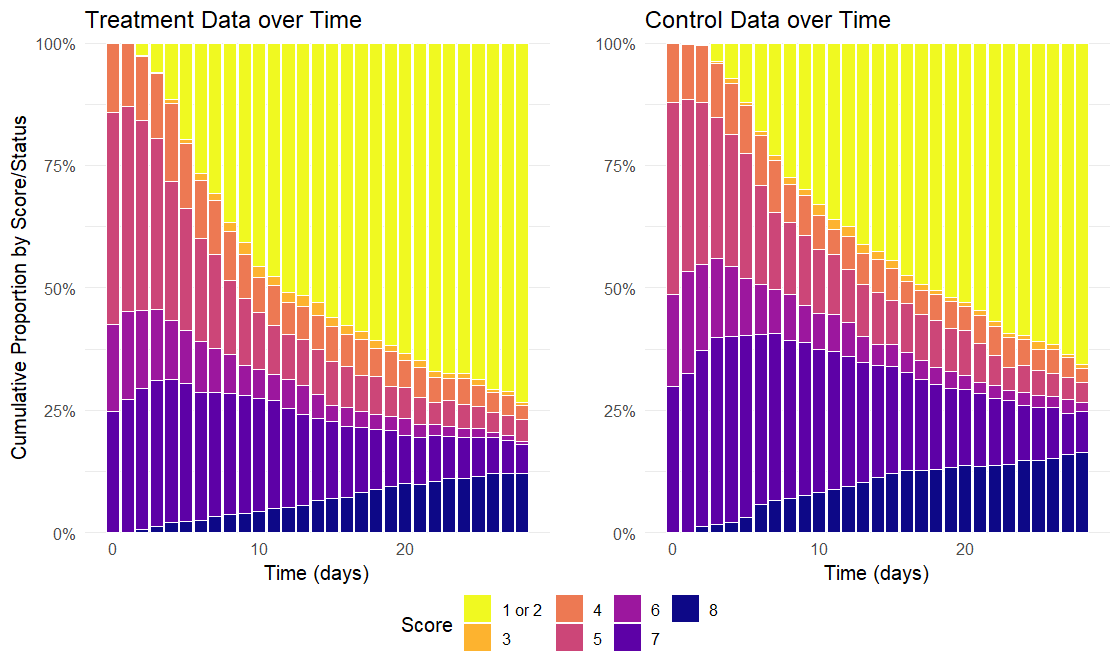}
   \begin{tablenotes}\tiny		
        \item The scores represent patients in different statuses as described by Table \ref{tab:scores}.
         A larger volume of not non-hospitalized scores in the treatment group (score of 1 or 2 in yellow) would indicate there is a treatment benefit, and thus a positive WINFT.
	\end{tablenotes}    \label{fig:actt_data_dist}
\end{figure}

Since a patient can be in any of the 7 states, neither RMT-IF nor EWT is applicable by lack of monotonicity. With our understanding of the data collection, we made the following additional data derivations on the daily severity score. The score on day 15 was a secondary endpoint, and it contained more information about the patient's state; we used it to fill in any missing scores between day 1 and day 15. We then used the recorded time-to-recovery to assign a score in category 1 or 2 to missing scores from the recovery time until the end of the study or until the patient was rehospitalized. If rehospitalized, any future missingness would be left as missing as there is no other available information to assign a score. The score is assigned to 8 at the time of death and any days afterwards. Any leftover missingness is handled by the estimation approach of WINFT by assuming data missing at random. With this data cleaning procedure, we had the distribution of daily scores for the treatment and control arms displayed in Figure \ref{fig:actt_data_dist}. 

\begin{figure}
    \centering
    \caption{ACTT-1 data: Proportion of wins, ties, and losses per day.}  
    \label{fig:ACTT1_Analysis}
    \includegraphics[trim={0.1cm 0.2cm 0.1cm 0cm},clip, width=0.9\linewidth]{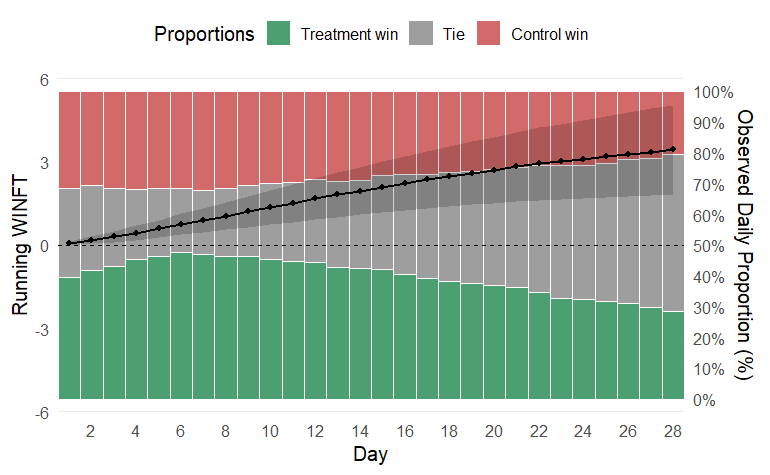}
      \begin{tablenotes}\tiny		
                 \item The proportion of pairwise wins, ties, and losses are plotted per day, in green, grey, and red, respectively (left axis).
                 \item The running WINFT (and associated 95\% pointwise confidence intervals) are plotted on each day in black (right axis).
			\end{tablenotes}
\end{figure}

Based on patients' states over time, it is clear that the volume of a 1 or 2 score (not hospitalized) is larger in the remdesivir group while its proportion of deaths over time appears to be lower. Both of these points suggest that remdesivir has some benefit over the placebo.
To estimate WINFT, for any pairwise comparison (remdesivir vs. placebo) on a given day, a higher score corresponds to a loss. We obtained a WINFT of 3.44 days with a 95\% confidence interval of 1.84-5.04 days of benefit of remdesivir over the placebo arm through the 28-day study duration. The calculations can be visualized in Figure \ref{fig:ACTT1_Analysis}, which displays the cumulative WINFT over time.


In Figure \ref{fig:ACTT1_Analysis}, the proportion of pairwise wins on any given day (Y-axis on the right) appears to be more than the proportion of pairwise losses, which contributes to the monotonically increasing WINFT estimate from day 1 through day 28. Thus, over the 28 days, there was a net 3.44 days benefit for remdesivir over placebo.

\subsection{HF-ACTION}

The Heart Failure: A Controlled Trial Investigating Outcomes of Exercise Training (HF-ACTION) was a multicenter randomized controlled trial conducted to evaluate the effects of exercise training on health status among patients with heart failure \citep{flynn_effects_2009}. In total, 2,331 patients were enrolled and randomized to either usual care (i.e., control) or usual care plus aerobic exercise (i.e., treatment). The study used the composite of all-cause mortality and hospitalizations as the primary endpoints and concluded that there was a statistically significant effect of exercise. 

For illustration, we use all cause mortality and the number of hospitalizations as the two longitudinal endpoints over the 4 year period, using the data available from the {\it rmt} R package \citep{mao_rmt_2021}. The data in the package include only a subgroup of 426 non-ischemic patients with a baseline cardio-pulmonary exercise test less than or equal to 9 minutes; 205 of the 426 patients were in the treatment group. Hospitalization and death (and associated times) are recorded for each participant as well as censoring time. Since hospitalizations were possibly recurrent, we define the first endpoint per day as the number of hospitalizations up until that day. The second, but more important in the hierarchy, endpoint is the daily indicator of death. Median follow up time was, respectively, 2.3 and 2.39 years for the treatment and control groups within this subgroup over this 4 year study.


\begin{figure}[ht]
    \centering
    \caption{HF-ACTION data: States of individuals still at risk through the duration of the trial.}
    \label{fig:EDA_HFACTION}
        \includegraphics[trim={0.2cm 0.4cm 0.2cm 0cm},clip, width=0.95\linewidth]{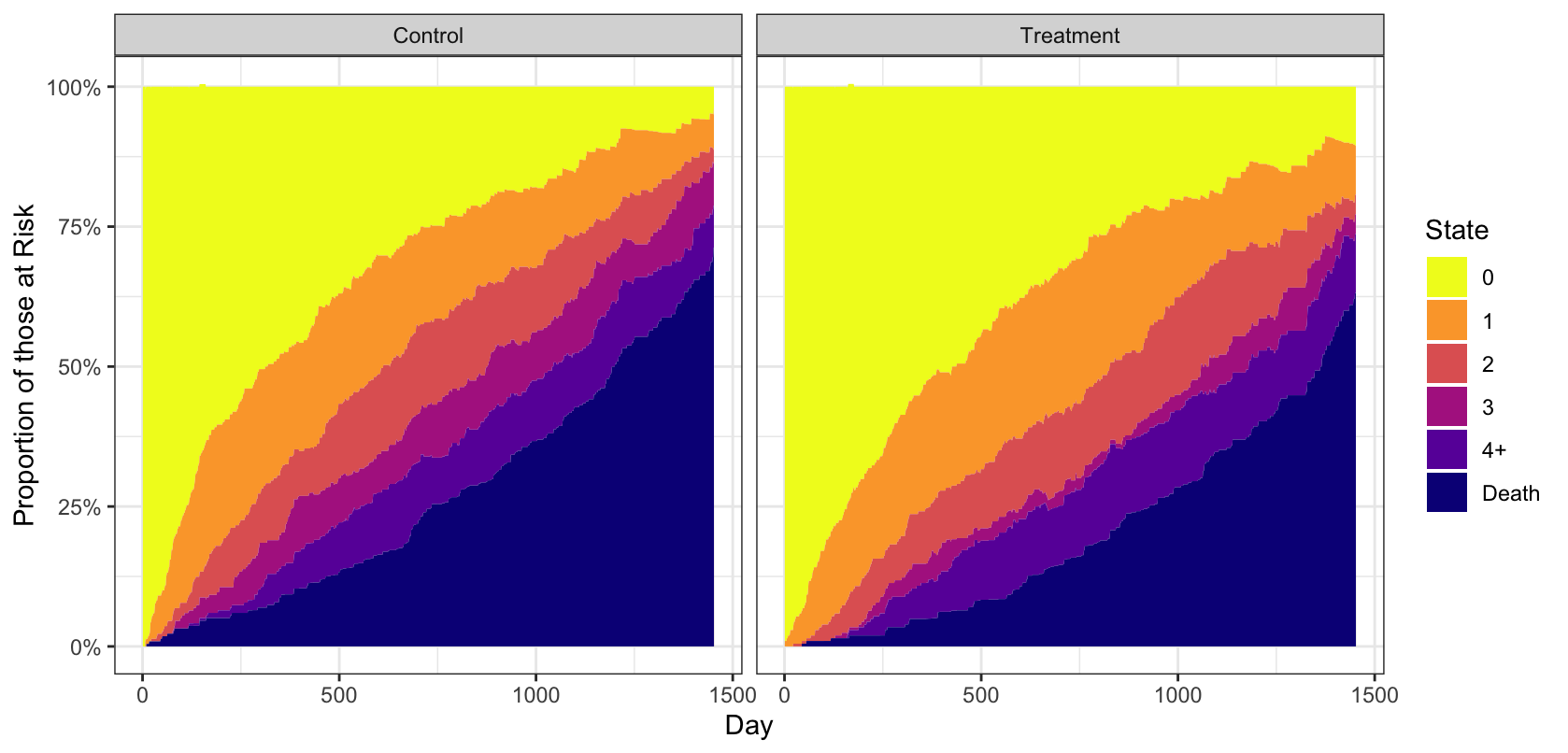}
    \begin{tablenotes}\tiny		
        \item States 0, 1, 2, 3, and 4+ denote the number of hospitalizations the patient had up until that time. 
        \item 
        The dark blue group represents the proportion (relative to those at risk) of those who were observed to have died.
	\end{tablenotes}
\end{figure}

As shown in Figure \ref{fig:EDA_HFACTION}, the smaller volume of the treatment group in the death state, along with a larger volume of patients without ever going to the hospital, suggest that there was a longitudinal benefit in adding exercise over the standard care.

\begin{figure}[h]
    \centering
    \caption{Daily proportions of treatment wins, losses, and ties along with the running WINFT and corresponding daily 95\% confidence interval.}
    \includegraphics[trim={0.2cm 0.2cm 0.1cm 0.4cm},clip, width=0.95\linewidth]{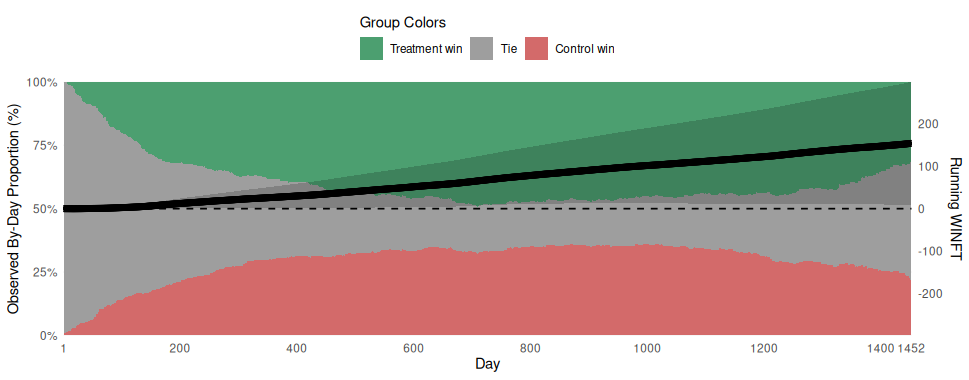}
    \label{fig:HF_Analysis}
\end{figure}

Running the actual WINFT procedure under this set up yields 153.98 days of benefit for the treatment over the control, with a 95\% confidence interval of (9.13, 298.83) days over the 4-year long study. This result is very close to the RMT-IF estimate of 5.05 months (i.e., 151 days) over the 4 years, reported by \cite{Mao_2023} based on the same data set. The 95\% confidence interval of (26, 277) days over the 4 years obtained by \cite{Mao_2023} is slightly shorter than the 95\% confidence interval for WINFT, which is expected based on our simulation study. Similar to the ACTT-1 example, we can calculate the proportions of wins, losses, and ties for all available pairwise comparisons by day and plot the running value of WINFT, as displayed in Figure \ref{fig:HF_Analysis}. 

Testing the null hypothesis
$H_0: \mathrm{WINFT} = 0 \text{ vs. } H_a: \mathrm{WINFT} > 0$
yields a p-value of 0.02, indicating that exercise  had a significant benefit, over the usual care, during the 4-year period. This p-value is nearly identical to the p-value of 0.019 derived from the RMT-IF procedure \citep{Mao_2023}. This result is expected as both WINFT and RMT-IF are estimating the same estimand in this scenario, but just use different approaches for inference.

\section{Discussion}\label{sec5:discussion}

We developed the {\it win time in favor of treatment} (WINFT) to provide a cumulative estimate of the net time that the treatment is in a better state than control. With this estimator, we generalize existing estimands for longitudinal hierachical endpoints, like RMT-IF and EWT, under one framework. The WINFT can handle non-monotonic endpoints and does not place any assumptions on the state space of the endpoints. This widely increases the number of applicable endpoints in a longitudinal analysis of the net winning of the treatment, from monotonic endpoints to endpoints like indicator functions for not being in hospital, or general non-monotone endpoints like quality-of-life endpoints.
We derived a closed-form variance estimator for WINFT using a $U$-statistic framework, which makes computation faster than other approaches that might require a bootstrap-like procedure. Finally, by weighting the uncensored wins or losses over time, the estimate for WINFT becomes robust under independent censoring, while inverse probability weighting deals with missing-completely-at-random data. 

The simulation results show that the proposed WINFT estimator is consistent for the true win time in favor for the treatment. Comparisons of variance estimates to Monte Carlo estimates show that the variance estimation via $U$-statistic approach provides good approximation. Moreover, we showed that WINFT, EWT and RMT-IF have the same estimands for time-to-event hierarchical endpoints under monotonicity requirements and with a pre-specified time horizon. While PWT also has the same estimand, its estimation approach does not include a mechanism to handle censoring and can yield different, and likely biased, estimator for data with censoring \citep{ozenne_asymptotic_2021}. The simulation studies also indicated that the WINFT has a larger variance due to the non-parametric nature, compared to the variance estimators for EWT and RMT-IF that utilize the monotonicity assumption in estimating state probability. While use of the monotonicity assumption may result in smaller variance estimates of EWT and RMT-IF, WINFT has the advantage of being nonparametric and applicable to any type of endpoint. Thus, while the power might be slightly lower for the WINFT in some scenarios, unlike the PWT, estimator of WINFT appears to be consistent in the presence of non-informative censoring and missing-completely-at-random data, and unlike RMT-IF and EWT, it requires no assumption of monotonicity or modeling of the state space.

Future work in WINFT includes covariate adjustment to get more precise estimates on the treatment effect or to account for potential bias due to covariate imbalance. Another question of interest is how to account for the uncertainty in estimating the survival function of time-to-censoring and missingness probabilities. Overall, as constructed, WINFT provides a generalized framework with nearly no assumptions for assessing the net time in favor of a treatment over a control based on any types of longitudinal hierarchical endpoints.  

\section*{Acknowledgements}
No acknowledgments.

\section*{Data Availability Statement}
The de-identified dataset from the ACTT-1 (NCT04280705) study was supplied by the NIAID Clinical Trials Data Repository, Accessclinicaldata@NIAID. The data can be requested from: https://accessclinicaldata.niaid.nih.gov/. The de-identified subgroup dataset from the HF-ACTION study can be accessed through the \textit{rmt} package \citep{mao_rmt_2021} in R.

\section*{Disclosure statement}
The authors have no conflicts of interest to disclose.

\section*{Use of Generative AI}
The authors used ChatGPT (OpenAI; GPT-5.4) for support with; programming, concept clarification, and literature review. All output was reviewed and verified by the authors.

\section*{Funding}
Research reported in this publication was supported by the National Heart, Lung, And Blood Institute of the National Institutes of Health under Award Number T32HL079896. The content is solely the responsibility of the authors and does not necessarily represent the official views of the National Institutes of Health. This research is also supported by the Duke Clinical Research Institute Biostatistics and Data Science Research Fund.




\bibliographystyle{plainnat}
\bibliography{bibliography.bib}

\newpage 
\appendix 
 
\section{Expectation and Variance of WINFT}\label{appA}
\subsection{Expectation}
Assume that both $G^z(t)$ and $o^z(t)$ are known and let $\mathcal{F}_{ij}^t:=\sigma\{D_i^A, D_j^B, Y_i^A(t), Y_j^B(t)\}$.
\begin{align*}
    \mathbb{E}[\hat\theta_t]&= \frac{1}{n_An_B}\sum_{i=1}^{n_A}\sum_{j=1}^{n_B}\mathbb{E}[\Omega_i^A(t)\Omega_j^B(t)W_{ij}]
\end{align*}

Then,
\begin{align*}
    \mathbb{E}[\Omega_i^A(t)\Omega_j^B(t)W_{ij}] &= \mathbb{E}\{\mathbb{E}[\Omega_i^A(t)\Omega_j^B(t)W_{ij}(t)\mid \mathcal{F}_{ij}^t]\}\\
    &= \mathbb{E}\{W_{ij}(t)\cdot \mathbb{E}[\Omega_i^A(t)\mid \mathcal{F}_{ij}^t]\cdot\mathbb{E}[\Omega_j^B(t)\mid \mathcal{F}_{ij}^t]\}
\end{align*}

Now we investigate $\mathbb{E}[\Omega_i^A(t)\mid \mathcal{F}_{ij}^t]$ by splitting it into the $A_i^A(t)$ and $B_i^A(t)$ segments (defined in \ref{estimation}).

\begin{enumerate}
    \item Expectation from $B_i^A(t)$ (death hasn't happened yet),
    \begin{align*}
        \mathbb{E}\left[\frac{B_i^z(t)}{G^z(t)o^z(t)}\mid \mathcal{F}_{ij}^t\right] &= \frac{\mathbb{I}\{D_i^z>t\}}{G^z(t)o^z(t)} \mathbb{E}[\mathbb{I}\{C_i^z>t\}O_i^z(t)\mid \mathcal{F}_{ij}^t] \\
        &= \frac{\mathbb{I}\{D_i^z>t\}}{G^z(t)o^z(t)} P(C_i^z>t, O_i^z(t)=1\mid \mathcal{F}_{ij}^t)\\
        &=  \frac{\mathbb{I}\{D_i^z>t\}}{G^z(t)o^z(t)} \underbrace{P(C_i^z>t\mid \mathcal{F}_{ij}^t)}_{=G^z(t)} \underbrace{P(O_i^z(t)=1 \mid \mathcal{F}_{ij}^t)}_{=o^z(t)}\\
        &= \mathbb{I}\{D_i^z>t\}
    \end{align*}

    \item Expectation from $A_i^A(t)$ (death observed),
    \begin{align*}
        \mathbb{E}\left[\frac{A_i^z(t)}{G^z(D_i^z)}\mid\mathcal{F}_{ij}^t\right] &= \frac{\mathbb{I}\{D_i^z<t\}}{G^z(D_i^z)}\mathbb{E}[\mathbb{I}\{D_i^z<C_i^z\}\mid \mathcal{F}_{ij}^t]\\
        &= \frac{\mathbb{I}\{D_i^z<t\}}{G^z(D_i^z)}P(D_i^z<C_i^z)\\
        &= \frac{\mathbb{I}\{D_i^z<t\}}{G^z(D_i^z)}G^z(D_i^z)\\
        &= \mathbb{I}\{D_i^z<t\}
    \end{align*}
\end{enumerate}

Then we combine the parts to see that:

\begin{align*}
    \mathbb{E}[\Omega_i^A(t)\mid \mathcal{F}_{ij}^t] &= \mathbb{E}\left[\frac{A_i^z(t)}{G^z(D_i^z)}\mid\mathcal{F}_{ij}^t\right] + \mathbb{E}\left[\frac{B_i^z(t)}{G^z(t)o^z(t)}\mid \mathcal{F}_{ij}^t\right] \\
    &= \mathbb{I}\{D_i^z<t\} + \mathbb{I}\{D_i^z\geq t\}\\
    &= 1
\end{align*}

Thus we have shown that $\mathbb{E}[\Omega_i^A(t)\mid \mathcal{F}_{ij}^t]$ and $\mathbb{E}[\Omega_i^B(t)\mid \mathcal{F}_{ij}^t]$ sum to 1 and therefore our estimator is unbiased. 

\subsection{Variance}
We can get the variance of WINFT through the Hajek projection principle for $U$-statistics. Start by defining $h_{1,0}(\bm{y}_i^A)$ and $h_{0,1}(\bm{y}_k^B)$ using $f$ defined earlier.

\begin{align*}
    h_{1,0}(\bm{y}_i^A) &:= \mathbb{E}\left[ f(\bm{y}_i^A,\bm{Y}_k^B) \right]-WINFT\\
    h_{0,1}(\bm{y}_k^B) &:= \mathbb{E}\left[ f(\bm{Y}_i^A,\bm{y}_k^B) \right]-WINFT
\end{align*}

Then the projection of $\widehat{WINFT}-WINFT$ is given by the following.

$$\frac{1}{n_A}\sum_{i=1}^{n_A}h_{1,0}(\bm{Y}_i^A) - \frac{1}{n_B}\sum_{k=1}^{n_B}h_{0,1}(\bm{Y}_k^B)$$

Then the variance of $\widehat{WINFT}$ then is the variance of the projection, yielding the following.

\begin{align*}
    \widehat{V}(\widehat{WINFT)} &= \frac{1}{n_A^2}\sum_{i=1}^{n_A}V(h_{1,0}(\bm{Y}_i^A)) - \frac{1}{n_B^2}\sum_{k=1}^{n_B}V(h_{0,1}(\bm{Y}_k^B))\\
    &= \frac{1}{n_A^2}\sum_{i=1}^{n_A}\mathbb{E}(h_{1,0}^2(\bm{Y}_i^A)) - \frac{1}{n_B^2}\sum_{k=1}^{n_B}\mathbb{E}(h_{0,1}^2(\bm{Y}_k^B))
\end{align*}

Where the first equality assumes that the covariance between $h_{1,0}$ and $h_{0,1}$ is negligible, and the second equality comes from the unbiasedness of the $f$ function for $WINFT$. With a single time point ($S=1$), this derivation closely follows the results of \cite{dong_generalized_2016}.

\section{ACTT-1 Simulation Study Setup}\label{appB}\medskip

We discuss here the simulation setup for validating WINFT from Section \ref{sec:valid_WINFT}. We start by using the data of Table 2 from the Remendisivir trial against COVID-19 \citep{beigel_remdesivir_2020}. This table outlines the number of participants who start in each state (1-8 from Table \ref{tab:scores}) in the treatment and control group, and the number of participants in each group at day 15. We then posit a Markov model assuming participants only change their state at most once during the trial, and that these state transition probabilities remain constant over the 15 days the patients are observed. We let $Y_t^{(g)}$ represent the distribution of scores of participants at time $t$ in arm $g$ (treatment or control). We assume that participants start in state $a$ with probability $q_a^{(g)}$. We then let the probability of transitioning from state $a$ to state $b$ be denoted with $p_{a,b}^{(g)}$. Note that all of these probabilities are arm-specific. With these definitions, we can then define the model as follows.
\begin{align}
   &P(Y_t^{(g)}=y_t|Y_{t-1}^{(g)}=y_{t-1}, Y_{t-2}^{(g)}=y_{t-2},...,Y_1^{(g)}=y_1) \\ 
   &:= 
\begin{cases}
    p_{(a,a)}^{(g)} \text{ if } y_{t}=y_{t-1}=y_{t-2}=...=y_{1} = a \text{ (no score change yet)} \\
    p_{(a,b)}^{(g)} \text{ if } b=y_t\neq y_{t-1} = y_{t-2} =...=y_1 =a\text{ (score change at time t)}\\
    1 \text{ if } y_{t}=y_{t-1}\neq y_1 \text{ (score already changed)}\\
    0 \text{ if } y_{t}\neq y_{t-1} \neq y_1 \text{ (score can't change twice)}
\end{cases}
\end{align}

We obtain the oracle initial state probabilities, $q_b^{(g)}$, by simply taking the proportion of participants that start in each group. Let $N_{b|a}^{(g)}$ denote the proportion of participants who start at score $A$ but end at score $B$ in arm $g$. Let $N_{\Sigma|A}^{(g)}$ denote the total number of participants that start at score $A$. Deriving the oracle transition probabilities then comes from solving the following two equations based on the observed proportion of participants within each group.
\begin{align*}
    \frac{N_{a|a}^{(g)}}{N_{\Sigma|a}^{(g)}} &= \left(p_{(a,a)}^{(g)}\right)^{14}\\
\frac{N_{b|a}^{(g)}}{N_{\Sigma|a}^{(g)}} &= p_{(a,b)}^{(g)}\sum_{k=0}^{13}\left(p_{(a,a)}^{(g)}\right)^k
\end{align*}

The sum appears as a consequence to the fact that the score can remain unchanged from anywhere between The code for the actual calculations can be found in the supplemental section. 

From the now-derived oracle transition probabilities, we can calculate the true WINFT assuming data collection on every day. We start by deriving the unconditional probability of being at a given state on an arbitrary day $t$, then use that formula to get the net beneift on that day, which is lastly summed over the 14 days. Day 1 is excluded as it was the patient's baseline score. The probability of being in a given state on day $t$ is found as follows.

\begin{align*}
    P(Y_t^{(g)} = b) &= \sum_{i=1}^8 P(Y_t^{(g)} = b, Y_1^{(g)}=i)\\
    &= \sum_{i=1}^8 P(Y_t^{(g)}=b|Y_1^{(g)}=i)P(Y_1^{(g)}=i)\\
    &= P(Y_t^{(g)}=b|Y_1^{(g)}=b)P(Y_1^{(g)}=b) + \sum_{i=1, i\neq b}^8 P(Y_t^{(g)}=b|Y_1^{(g)}=i)P(Y_1^{(g)}=1)\\
    &= \left(p_{(b,b)}^{(g)}\right)^{t-1}q^{(g)}_b + \sum_{i=1, i\neq b}^8 q_b^{(g)}\left\{p_{(i,b)}^{(g)}\sum_{j=1}^{t-2}\left(p_{(i,i)}^{(g)}\right)^j\right\}
\end{align*}

Where the sum is induced for the same reasons as in the derivation of the oracle probabilities. Now equipped with the probabilities of the treatment and control being in a state $b$ at time $t$, calculating daily net benefit is straightforward. If we let $g=1$ and $g=0$ imply the treatment and control group respectively, the net benefit on a given day by $P\left(Y_t^{(1)} < Y_t^{(0)} \right) - P\left(Y_t^{(1)} > Y_t^{(0)} \right)$ as a higher score is worse according to Table \ref{tab:scores}. By calculating these probabilities according to this procedure, then summing for each day to calculate the true WINFT, we get 1.77 days net benefit of the treatment over the control over the 15 day period. We can then simulate data using the Markov model to analyze the estimation of WINFT.

\section{Simulation Procedure for Comparative Analysis}
\label{appE}

This method follows from \cite{troendle_use_2024} and their adaptation of methods from \cite{beyersmann_simulating_2009}. We start by assuming a 1:1 randomization of patients to treatment and control. We then Let $Z=1$ for patients in the active treatment and $Z=0$ for patients not. We then posit two competing risks at time $t_j$. The first is that the participant dies, and the second is that the participant is lost to follow up or the study terminates at time $t_j$. We then order the events from least important to most; myocardial infarction (MI) $<$ heart failure (HF) $<$ death. We then set a cause specific hazard model for the four events (the 3 outcomes and censoring) that are proportional on the treatment indicator

\begin{align*}
    \text{MI: } \lambda_1(t)&=\exp\{\alpha_1+\beta_1Z + Y\}\\
    \text{HF: } \lambda_2(t)&=\exp\{\alpha_2+\beta_2Z + Y\}\\
    \text{Death: } \lambda_3(t)&=\exp\{\alpha_3+\beta_3Z + Y\}\\
    \text{Censoring: } \lambda_{4}(t)&=\exp\{\alpha_4\}
\end{align*}
where $Y\sim U(-0.25, 0.25)$ is the frailty variable. 

For a given a person, we then simulate the time to first even from $\lambda=\displaystyle \sum_{i=1}^4\lambda_i(t)$, and the event is decided by a categorical distribution where the probability of event $i$ is $\displaystyle \frac{\lambda_i}{\lambda}$. The chosen even is removed from the total probability and the second time is calculated and so on. Similar to \cite{troendle_use_2024} we used a 1 year administrative cutoff. Following this schema, we used $\alpha_1=\alpha_2=0.1, \alpha_3=-1.2$.

\end{document}